\documentclass[12pt, a4paper]{article}
\usepackage{authblk}
\usepackage{hyperref}
\usepackage{soul}
\usepackage{longtable}
\usepackage{comment}
\usepackage{multirow}
\usepackage{lineno}
\usepackage{setspace}

\usepackage[utf8]{inputenc}
\usepackage[ngerman, english]{babel} 
\usepackage{eurosym} 
\usepackage{amsmath}
\usepackage{tabularx} 
\usepackage{threeparttable} 
\usepackage{multirow} 
\usepackage{booktabs} 
\usepackage{pdflscape}
\usepackage{lscape} 
\usepackage{longtable}
\usepackage{dcolumn}
\usepackage[rounding]{rccol} 
\usepackage{morefloats} 
\usepackage{array}
\usepackage{fltpoint}[2004/11/12]
\usepackage{dcolumn}
\usepackage{rccol}
\usepackage{numprint}
\usepackage{listings}
\usepackage{csquotes}
\usepackage{lscape}
\usepackage{longtable} 
\usepackage[T1]{fontenc}
\usepackage{adjustbox}
\usepackage{numprint}
\usepackage{ulem}

\usepackage[font=small]{caption}
\usepackage{subcaption}
\usepackage{dsfont}

\usepackage{listings} 
\usepackage[authoryear]{natbib}

\usepackage{amsthm}
\usepackage{amsmath}
\usepackage{amsbsy}
\usepackage{amssymb}
\usepackage{comment} 

\usepackage[left=2.5cm, top=2.5cm, bottom=2.5cm, right=2.5cm]{geometry} 
\usepackage{tocloft} 
\usepackage{hyperref}
\usepackage{yhmath}
\usepackage{rotating}
\usepackage{graphicx}
\usepackage{textcomp}
\usepackage{mathtools}

\usepackage{tabularx}
\usepackage{tikz}
\usetikzlibrary{arrows}

\usetikzlibrary{decorations,arrows}
\usetikzlibrary{decorations.pathmorphing}
\usepgflibrary{decorations.pathreplacing} 
\usetikzlibrary{decorations.fractals}
\usetikzlibrary{decorations.footprints}
\usetikzlibrary{decorations.shapes}
\usetikzlibrary{decorations.text}
\usetikzlibrary{decorations.pathmorphing}
\usetikzlibrary{patterns}
\usepackage{epsfig}
 \usepackage{pst-grad} 
 \usepackage{pst-plot} 
\usepackage{filecontents}

\usepackage[font=small, labelfont=bf, justification=centering, singlelinecheck=false]{caption}

\renewcommand\appendix{\par%
\setcounter{section}{0}%
\setcounter{figure}{0}%
\setcounter{table}{0}%
\renewcommand\thesection{\Alph{section}}%
\renewcommand\thetable{\Alph{section}.\arabic{table}}%
\renewcommand\thefigure{\Alph{section}.\arabic{figure}}} %

\newcolumntype{M}{>{\centering\arraybackslash}m{1cm}}

\makeatletter
\def\blfootnote{\xdef\@thefnmark{}\@footnotetext} 
\makeatother

\defcitealias{SVR2006}{SVR (2006)}
\addtocontents{toc}{\protect\thispagestyle{empty}}

\graphicspath{ {./figures/} }

\title{Towards Measuring Disruptive Innovation Across Countries}
\author{
  Christian Rutzer$^{a,*}$%
  \quad
  Dragan Filimonovic$^{a}$%
  \quad
  Jeffrey T. Macher$^{b}$%
  \quad
  Rolf Weder$^{a}$%
}

\date{March 18, 2026}

\begingroup

\footnotetext{\hspace{-1.8em}
  \hangindent=1.2em \hangafter=1 \noindent
  $^{a}$\,Faculty of Business and Economics, University of Basel,
  Peter Merian-Weg 6, 4002 Basel, Switzerland.\par
  \hangindent=1.2em \hangafter=1 \noindent
  $^{b}$\,Robert E. McDonough School of Business, Georgetown University,
  37th and O Streets NW, Washington, DC 20057, USA.\par
  \hangindent=1.2em \hangafter=1 \noindent
  $^{*}$\,Corresponding author: christian.rutzer@unibas.ch%
}
\endgroup

\date{\today}

\usepackage{footmisc}

\begin{document}
\maketitle
\begin{abstract} 
\noindent The $CD$ index is a widely used measure of disruptive inventions. Most studies compute it using USPTO data. This creates a puzzle because the US appears less disruptive than European and Asian countries. We show that this largely stems from missing international citations. Using a global citation network, we quantify and correct this bias. The disruptiveness advantage of non-US inventors drops by 64\% to 148\% of the US baseline mean. The US emerges as a disruption leader over Europe, with Asia's advantage substantially reduced. Globally integrated citation data are essential for credible measurement of disruptive innovation in international contexts.
\end{abstract}


\vspace*{0.5cm}

\newpage 

\section{Introduction}

Which countries lead in producing disruptive innovations that make prior technologies obsolete rather than merely extending them? As global competition for technological leadership intensifies, answering this question is a first-order concern for both policymakers and scholars of innovation and economic growth \citep{Schumpeter1941, Rosenberg1982inside, Aghion1992, Christensen1997,furman2002}. Yet, measuring disruptive innovation across countries has proven difficult. Standard patent metrics capture the impact or novelty of inventive output but not whether an invention displaces prior technologies.\footnote{See, e.g., \citet{trajtenberg1990penny} and \citet{hall2001, hall_2005} for citation-based measures and, e.g., \citet{Kelly2021} and \citet{arts2021, arts2025} for text-based approaches. \citet{yang2025text} compares the $CD$ index with the text-based KI index of \citet{Kelly2021}.} 

The Consolidation-Disruption ($CD$) index introduced by \citet{funk2017dynamic} helps to fill this gap by providing a structural measure of Schumpeterian creative destruction in patents based on citation network topology. It distinguishes between patents that render prior work obsolete (disruptive) and those that consolidate existing knowledge streams (consolidating), depending on the citation structure of follow-on patents. The $CD$ index has been validated in both scientific and patent domains \citep{funk2017dynamic, wu2019}, and has become a standard metric in high-profile studies of innovation dynamics \citep[e.g.,][]{park2023papers, frey_2023}.

A striking paradox emerges, however, when the $CD$ index is constructed using US Patent and Trademark Office (USPTO) data (following standard research practice): in particular, the United States (US) trails Japan, South Korea, China, and every major European country in disruptive innovation. This outcome starkly contradicts the extensive body of evidence documenting US technological leadership \citep[e.g.,][]{van2008productivity, aghion2021power, WIPO2025}.

We show that this apparent US innovation deficit is a direct result of measurement bias. The validity of the $CD$ index rests on a critical, yet often implicit, assumption that the citation network is complete. This assumption breaks down when citation networks are restricted to a single jurisdiction (e.g., USPTO), causing inventions to lose observable links to foreign prior art. Because inventors and examiners exhibit a well-documented citation home bias \citep{jaffe1993geographic, jaffe_2017}, this truncation is not random. Missing citations disproportionately affect patents originating from foreign inventors, artificially inflating their measured disruptiveness. The result is a "disruption mirage", in which the apparent disruptiveness advantage of non-US inventors largely reflects data incompleteness rather than genuine differences in innovative capacity.

We formalize this identification failure and quantify its empirical importance by constructing a globally integrated patent citation dataset based on the patent statistics (PATSTAT) of the European Patent Office (EPO). We retain USPTO patents as the focal set to ensure legal and economic comparability and incorporate international prior art to recover missing technological lineages. Using a stacked regression framework that simultaneously identifies structural country differences and source-specific measurement bias, we find that single-jurisdiction citation data inflates the measured disruptiveness of non-US inventors relative to US inventors by 64\% to 148\% of the US baseline mean. In the corrected data, US patents are significantly more disruptive on average than those from every major European country, and the massive apparent advantage of Japan, South Korea, and China shrinks substantially. Incorporating global citation data is thus not merely a refinement but a prerequisite for valid cross-country comparisons of innovative capacity.

Building on this framework, we contribute to four research streams. First, we add to the literature that uses the $CD$ index to examine disruptive inventions. Prior studies have applied the $CD$ index to various drivers of disruptiveness, including organizational settings \citep{funk2017dynamic}, team size \citep{wu2019}, inventor age \citep{kaltenberg2023invention}, geography \citep{frey_2023}, and long-term trends \cite[]{park2023papers,macher2024,yang2025text,park2025,wu2026innovation}. This research has primarily relied on USPTO data and focused on aggregate trends pooled across all inventors, with limited insight into national differences. We offer the first systematic \textit{country-level comparison}, enabled by extending the citation network beyond USPTO data to incorporate foreign prior art. 

Second, we contribute to the literature on measurement error in innovation metrics. We follow the tradition of identifying systematic biases in patent data construction, similar to \citet{alcacer2006patent} for examiner citations, and apply this lens to the $CD$ index. Most existing critiques have considered scientific citation networks, addressing computational sensitivities \citep{wei2023quantifying, petersen2023disruption, leibel2024we} or bibliographic coverage issues such as older work or book underrepresentation \citep{ruan2021rethinking, liang2022revisiting, holst2024dataset, park2025}. We identify a structural bias specific to patent citation networks: namely, the mismatch between global knowledge flows and jurisdictional data boundaries. Unlike scientific publications which theoretically form a single global corpus, patent data are inherently fragmented by national legal systems. We show that relying exclusively on, e.g., USPTO data introduces a specific "jurisdictional boundary" bias. We build on \citet{macher2024}, who document how changes in USPTO citation practice bias the $CD$ index over time, and show that cross-patent office citation omissions constitute an additional and distinct bias source.  

Third, we add to the methodological literature on international innovation comparisons. Prior scholarship has identified how data matching choices \citep{thompson2006patent} and legal institutions \citep{moser2005} bias standard citation volume and knowledge flow metrics. Researchers have addressed these issues through harmonized patent data via PCT applications and triadic patent families \citep{boeing2016,de2013worldwide,squicciarini2013} or examiner versus inventor citation decomposition \citep{criscuolo2008}. We extend this methodological rigor to the $CD$ index, where the consequences of missing data are qualitatively different. Missing cross-border citations not only decrease measured knowledge flows, but also systematically inflate disruptiveness scores, both by omitting follow-on patents from foreign offices and by obscuring links between observed follow-on patents and foreign prior art. Therefore, global citation integration is a prerequisite for reliable international comparisons of $CD$ index values.

Fourth, we contribute to the substantive literature on national innovation capacity \citep{furman2002, jang2016, raghupathi2017innovation}. We advance recent efforts to move beyond simple patent counts toward quality-adjusted innovation metrics---such as stock-market valuations \citep{kogan2017} or text-based measures \citep{Kelly2021}---by providing what is, to our knowledge, the first disruptiveness-based ranking of national innovation systems. By bridging network-based metrics with national benchmarking, we offer new insights into which nations lead in redefining technological frontiers, a distinction of first-order importance for long-term economic growth.


\section{The $CD$ Index Measurement Puzzle}

We measure disruptive innovation using the $CD$ index introduced by \cite{funk2017dynamic}. Although the literature has proposed several variants \citep{leibel2024we}, we rely on the original and most widely used specification. Adopting the notation from \cite{wu2019}, the $CD$ index for a focal patent $i$ is defined as:

\begin{align}\label{cd_index}
CD_{i} = \frac{N_F - N_C}{N_F + N_P + N_C}.
\end{align}

The $CD$ index incorporates three distinct forward citation count measures: Focal-only ($N_F$) that reference the focal patent but none of its predecessors; Predecessor-only ($N_P$) that reference only the predecessor patents of the focal patent; and Combined ($N_C$) that reference both the focal patent and at least one of its predecessors.

The $CD$ index ranges from $-1$ (fully consolidating) to $1$ (fully disruptive) and positions a focal patent within the flow of subsequent inventive knowledge. A value of $1$ indicates that all subsequent inventions cite the focal patent without citing its predecessors, whereas a value of $-1$ indicates that all subsequent inventions cite both the focal patent and at least one of its predecessors. Intermediate values reflect the relative tendency of follow-on inventions to rely more on the focal patent or on its predecessors. Because the index is constructed entirely from observed citation links, incomplete citation coverage can mechanically distort these counts and thus bias measurement.

\begin{figure}[!ht]
\caption{Average Disruptiveness by Country using USPTO Data}
\label{fig:intro_uspto_bias}
\centering
\captionsetup{justification=centering}
\includegraphics[width=0.6\linewidth,keepaspectratio]{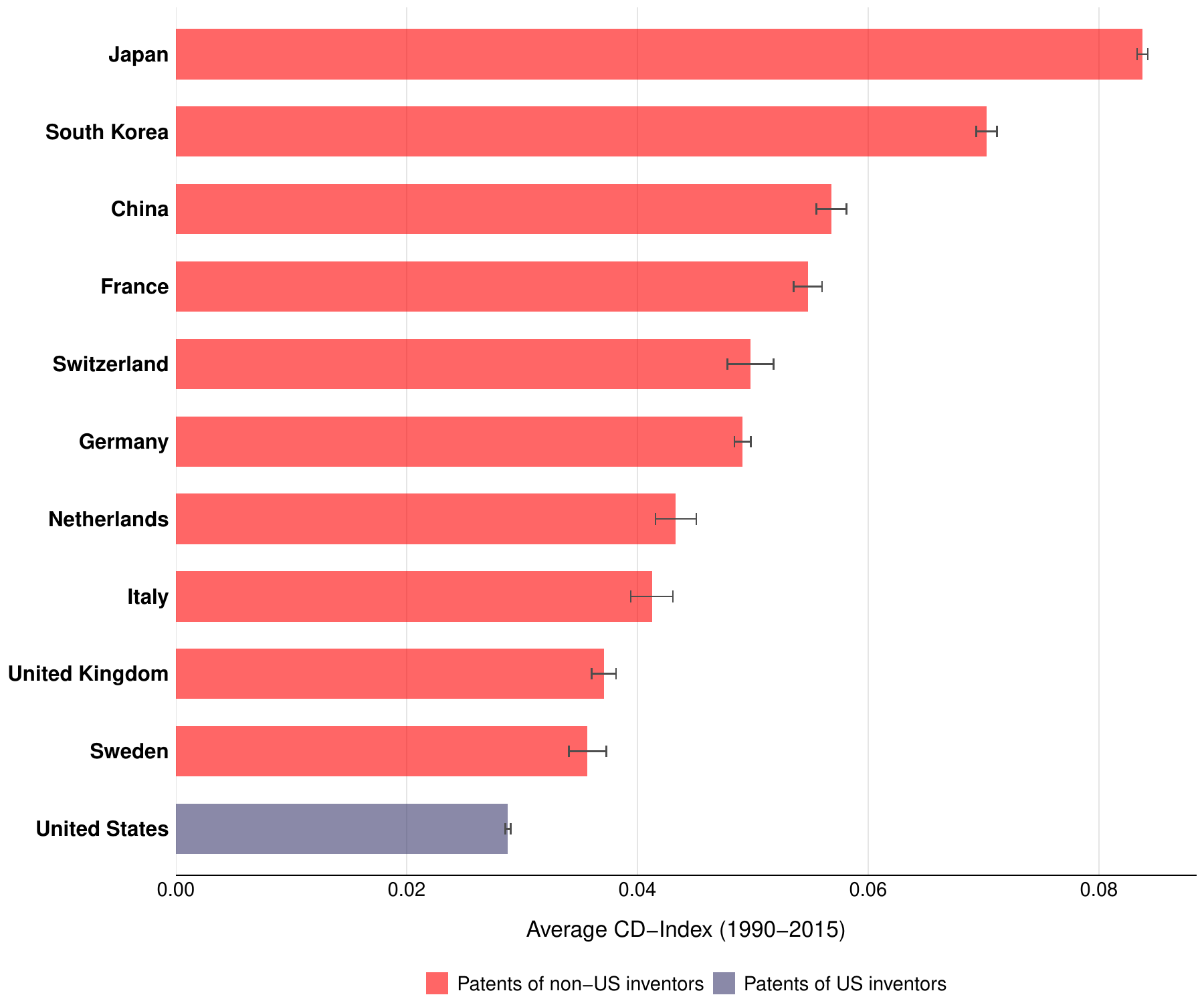}\par\vspace{0.5cm}
\begin{minipage}[b]{\textwidth}
    \footnotesize
    \emph{Notes:} The figure displays the average $CD$ index for USPTO patents by inventors from leading countries, using USPTO citation data only. $CD$ index measurement follows \citet{funk2017dynamic} and is at the patent family level. The sample is pooled for earliest publication years from 1990 to 2015. Error bars represent 95\% confidence intervals. Appendix Figure \ref{fig:app_single_patent} verifies the observed discrepancies using standard single-patent counts. Despite commonly perceived US aggregate technological leadership, US inventions exhibit significantly lower average disruptiveness than other major inventor countries.
\end{minipage}
\end{figure}

We start by computing the CD index following standard practice in the literature, using USPTO citation data only \citep[e.g.,][]{funk2017dynamic, park2023papers}. To ensure strict comparability with our subsequent analysis, we deduplicate USPTO records at the DOCDB patent family level, i.e., groups of related patents that cover the same technical content. Figure~\ref{fig:intro_uspto_bias} shows average $CD$ indices by country. A surprising puzzle emerges: patents by US inventors appear systematically less disruptive than patents by inventors from other leading innovator countries in Europe and East Asia. The magnitude of these differences suggests a substantial US disruptive innovation deficit relative to Europe and especially Japan and South Korea. This ranking stands in sharp contrast to the extensive evidence documenting US technological leadership \citep[e.g.,][]{van2008productivity, aghion2021power, WIPO2025}, raising the question of whether the observed pattern reflects genuine innovation differences or a measurement artifact.

We show that standard $CD$ index calculations introduce systematic measurement distortions in international comparisons. Because inventors and examiners typically exhibit a citation "home bias" \citep{jaffe1993geographic, jaffe_2017}, reliance on single-patent office data disproportionately obscures the citation histories of foreign inventions. To assess the extent to which these asymmetries explain the observed patterns, we develop a framework that accounts for the globally-integrated structure of patent citations.

\section{Analytical Framework and Data}\label{sec:cd_index}

This section formally defines citation truncation and its impact on the $CD$ index. We first derive the theoretical bias at the patent and country levels before detailing the data and empirical strategy used to quantify these distortions in practice.

\subsection*{Patent- and Country-Level Bias Derivation}

To formally derive how data truncation affects the $CD$ index in equation (\ref{cd_index}), we compare two citation samples for the same focal patent: a restricted (e.g., USPTO) sample ($s=1$) and an extended (e.g., augmented or complete) sample ($s=0$). The patent-level citation bias ($\Delta CD_i$) is defined as:

\begin{align}\label{cd_index_discrete}\Delta CD_{i} = CD_{i,1} - CD_{i,0} = \frac{N_{F,1}-N_{C,1}}{N_{F,1}+N_{P,1}+N_{C,1}} - \frac{N_{F,0}-N_{C,0}}{N_{F,0}+N_{P,0}+N_{C,0}}.
\end{align}

Equation~\eqref{cd_index_discrete} illustrates how structural citation network differences determine the bias. As the restricted sample ($s=1$) is a subset of the extended sample ($s=0$), some citing patents and specific citation links are systematically omitted. Because combined citations ($N_C$) require observing both focal and predecessor links, truncation implies that $N_{C,1} \leq N_{C,0}$. The effects of truncation on focal-only ($N_F$) and predecessor-only ($N_P$) citations are, however, ambiguous.\footnote{For notational simplicity, we omit the index $i$ from the component terms ($N_{F,s}, N_{P,s}, N_{C,s}$) but recognize that these counts are patent-specific.} 

While patent citation omissions tend to reduce baseline citation counts, specific citation link omissions can mechanically reclassify remaining citations: e.g., missing predecessor links shift counts from $N_C$ to $N_F$; missing focal links (with predecessor links intact) can shift counts from $N_C$ to $N_P$. It is consequently possible that $N_{F,1} > N_{F,0}$ or $N_{P,1} > N_{P,0}$ despite the smaller sample size. More generally, two mechanisms drive systematic bias in the $CD$ index: omissions and misclassifications. Omissions affect individual focal ($N_F$), combined ($N_C$) or predecessor ($N_P$) citations and thus change the denominator ($N_P$) or both the numerator and the denominator ($N_F$, $N_C$). Misclassifications (e.g., $N_C \to N_F$ or $N_C \to N_P$) leave the denominator intact but strictly increase the numerator, thereby amplifying the bias. Because these citation types enter the numerator and denominator asymmetrically in relative terms, the sign of the resulting patent-level bias in equation~\eqref{cd_index_discrete} is theoretically indeterminate. Data truncation can therefore affect the $CD$ index in either direction, which makes the net direction of the bias an empirical question.

For patents with positive $CD$ values, predecessor ($N_P$) and combined ($N_C$) citation omissions inflate the $CD$ index. Assuming pure omissions dominate misclassifications: a smaller $N_P$ decreases the denominator; a smaller $N_C$ has a compounding effect by simultaneously decreasing the denominator and increasing the numerator. By contrast, focal-only citation omissions ($N_{F,1} < N_{F,0}$) reduce measured disruptiveness. Misclassifications instead \textit{magnify} the overall bias. When a combined citation is recorded as focal-only ($C\!\to\!F$) in the restricted sample due to a missing predecessor link, the numerator increases but the denominator remains constant which increases measured disruptiveness.\footnote{Consider a citation misclassification recorded as focal-only in the restricted sample ($s=1$) but is combined in the extended sample ($s=0$) due to a missing predecessor link. This implies $N_{F,1} = N_{F,0} + \Delta$ and $N_{C,1} = N_{C,0} - \Delta$ for some positive integer shift $\Delta$ (e.g., $\Delta=1$). Calculating the bias $\Delta CD_i = CD_{i,1} - CD_{i,0}$ yields a strictly positive result, as the restricted sample measure $CD_{i,1}$ assigns disruptive weight to what is actually a combined link.} Crucially, for any given number of misclassified citations ($\Delta$), an $N_C \to N_P$ misclassification leaves the denominator unchanged and increases the numerator by only $\Delta$. In contrast, an $N_C \to N_F$ misclassification increases the numerator by $2\Delta$. Because this $C \to F$ shift exerts exactly twice the inflationary impact per occurrence, it acts as a potent mathematical multiplier, disproportionately amplifying the disruption bias whenever technological lineages are truncated.

For patents with negative $CD$ values, the net effect of citation omissions is also ambiguous. Pure focal-only ($N_F$) or predecessor-only ($N_P$) citation omissions in the restricted sample decrease the $CD$ index. The latter occurs because dividing a negative numerator by a smaller positive denominator results in a larger absolute, but more negative, value (assuming pure $N_P$ omissions outnumber $N_C \to N_P$ misclassifications, resulting in $N_{P,1} < N_{P,0}$). Conversely, combined ($N_C$) citation omissions increase the $CD$ index. Combined citation misclassifications ($C\!\to\!F$ and $C\!\to\!P$) increase the $CD$ index by strictly increasing the numerator while leaving the denominator constant.

A comparison of the positive and negative regions of the $CD$ index reveals asymmetric effects from missing background citations ($N_P$, $N_C$). For positive values, pure $N_P$ or $N_C$ omissions systematically compound to increase the $CD$ index. For negative values, pure omissions partially offset each other: missing $N_P$ links decrease the $CD$ index (i.e., making values more negative), while missing $N_C$ links increase the $CD$ index (i.e., making values less negative). The bias magnitude is consequently expected to be larger and more systematically positive in the positive region than in the negative region, \textit{ceteris paribus}. 

We quantify how these patent-level distortions translate to the country-level by applying this framework to the restricted ($s=1$) and extended sample data ($s=0$). For each country~$c$, we denote the expected disruptiveness of patents when measured using citation data $s \in \{0,1\}$ as:

\begin{equation} 
m_{c,s} = \mathbb{E}\!\left[\, CD_{i,s} \mid \text{ctry}_i = c \,\right]
\end{equation}

\noindent Note that $m_{c,1}$ corresponds directly to the (average) $CD$ index of an individual country ($c$) shown in Figure~\ref{fig:intro_uspto_bias} using USPTO data ($s=1$). We then define the country-level bias as:
\begin{equation} 
\text{Bias}_c = \mathbb{E}[\Delta CD_i \mid \text{ctry}_i = c] = m_{c,1} - m_{c,0}
\end{equation}

A positive (negative) value of $\text{Bias}_c$ indicates that the restricted citation coverage systematically inflates (deflates) measured patent disruptiveness for country~$c$.\footnote{The asymmetric missing citation effects imply that the country-level bias is not uniform. Missing background citations are likely to inflate $CD$ values more strongly in the positive range than in the negative range, where the effects of different omission types partially offset. Countries whose patent portfolios contain a larger share of genuinely disruptive inventions are therefore likely to experience a disproportionately larger upward bias in the restricted data.} To compare countries to a common benchmark, we define the \textit{relative bias} of country~$c$ with respect to the US as:
\begin{equation}\label{eq:relbias_def} 
\text{RelBias}_c = \text{Bias}_c - \text{Bias}_{\mathrm{US}} = (m_{c,1} - m_{c,0}) - (m_{\mathrm{US},1} - m_{\mathrm{US},0}).
\end{equation}

A positive (negative) $\text{RelBias}_c$ value implies that restricted citation coverage disproportionately inflates (deflates) measured disruptiveness for country~$c$, relative to the US.

\subsection*{Data}\label{sec:data}

We empirically evaluate the theoretically-derived bias above using two citation datasets that cover the same underlying inventions but differ in the observed citation network. The first is a USPTO benchmark based on PatentsView (release of September 10, 2025) from which we retain utility patents only. The second is an extension that augments these USPTO patents with citations recorded by all patent offices included in PATSTAT (Spring 2025 edition).\footnote{The PATSTAT database maintained by the EPO provides bibliographic and citation data for patents filed worldwide \citep{patstat}. PatentsView records citations made by USPTO examiners and applicants, which can include references to foreign patents. It does not capture citations made at foreign patent offices, however, meaning that non-USPTO forward citations are unobserved. We use only USPTO-recorded citation links in our restricted sample, consistent with standard practice in the literature.} We treat the extended PATSTAT citation network as more complete in measurement. The only dimension where the two samples differ is the set of observed citation edges.

Our unit of analysis is the DOCDB patent family, which we treat as a unique invention identifier. This deduplication is essential because national patent offices often cite different documents for the same underlying prior art (e.g., a US patent vs. an EPO equivalent). Mapping patents to families allows us to integrate these disparate links, eliminate office-specific citation practices, and avoid double-counting. Crucially, this provides the common identifier necessary to ensure strict comparability between the restricted USPTO and extended PATSTAT networks, which is a structural requirement for computing the patent-level bias $\Delta CD_i$ in Equation \ref{cd_index_discrete}. Because PatentsView does not contain international family identifiers, we use PATSTAT to assign DOCDB family codes to all USPTO patents. We note that PATSTAT is used in the restricted sample solely as a concordance table for family assignment. All citation links in the restricted network are derived exclusively from PatentsView.

We restrict the sample to focal patent families containing at least one USPTO utility patent for two purposes. First, it ensures comparability with the established $CD$ index literature \citep{funk2017dynamic, wu2019, park2023papers}. Second, and more importantly, it ensures that every focal invention has been examined and granted under a common institutional regime. This shared USPTO touchpoint means that all inventions in our sample are subject to the same patentability standards and disclosure requirements at least at one office, reducing the risk that observed cross-country differences reflect institutional heterogeneity rather than citation network structural properties.

Each patent family is assigned to technology fields using the WIPO one-digit technology classification, and is attributed to inventor countries based on PATSTAT-indicated inventor location(s). When inventor location is missing from the original filing, we impute it from other patent family members following \cite{Rassenfosse2019}. A single invention can be linked to multiple countries. 

Patent families are dated by their earliest worldwide publication year, regardless of whether it is an application or a grant. We follow standard research practice in our baseline estimation by counting forward citations within a post-publication five-year window, as a fixed horizon facilitates comparability across patents from different years. Section \ref{sec_robust} shows that the results are robust to using the earliest worldwide filing year as the reference date or extending to a ten-year forward citation window. For the restricted USPTO data, we retain only citations where both the citing and the cited families include at least one USPTO grant.\footnote{By focusing on patent families, our approach also consolidates application and grant citations, consistent with \cite{macher2024}.} For the extended PATSTAT data, we do not require that the citing family or any of the cited families realize a USPTO application or grant, since many valid citations originate from foreign jurisdictions. We do however require that each family results in a granted patent in at least one jurisdiction to ensure that the invention meets patentability standards somewhere. These conventions together provide a common time anchor and incorporate international patent citations not captured in USPTO-only data, thereby enabling more meaningful cross-country comparisons of patent disruptiveness.

Relying on international PATSTAT data nevertheless introduces dual temporal truncation threats: (1) incomplete backward citation record transfers from foreign patent offices to PATSTAT in earlier years; and (2) database ingestion lags in recent years. Both threats are consequential for computing the $CD_5$ index, which ideally requires complete backward citation records and a five-year forward citation window for each focal patent. We quantify structural missingness in the PATSTAT citation network by constructing a family-level coverage metric that reflects each patent family's vulnerability to incomplete records at its most relevant patent office, proceeding in three steps.

First, we link each patent family to its inventors' countries and to the set of patent offices at which it holds publications. For each patent office and filing year, PATSTAT provides official coverage statistics reporting the share of applications with available backward citation data \citep{patstat2025coverage}. We use these statistics to assign a time-varying coverage rate to every family-office pair. Second, we identify for each patent family the "Regional Home Office", defined as the patent office of the inventor's country of residence. This office is most likely to process the bulk of domestic prior art citations. For inventors based in European Patent Convention member states, we treat EPO filings as equivalent to home-office filings. The home office coverage rate constitutes the binding constraint on backward citation completeness for the family, since missing home office backward citations cannot be recovered from filings elsewhere. For the minority of families that bypass their home office entirely, we take the maximum coverage rate across all offices at which the family was published. Third, we aggregate the resulting family-level coverage rates to the inventor-country-publication-year level, requiring cohorts at this level contain at least 1,000 families to ensure stable estimates.\footnote{The coverage rate encompasses all patent families (not only focal USPTO families) published at a given office-year because the backward citation records of both focal inventions and citing patents, as well as the proper identification of cited predecessors in the database, affect the citation structure completeness underlying the $CD$ index.} 

\begin{figure}[!ht]\caption{Citation Coverage and Sample Selection Window.}\label{pat_coverage}
\centering\captionsetup{justification=centering}
\includegraphics[width=0.9\linewidth,keepaspectratio]{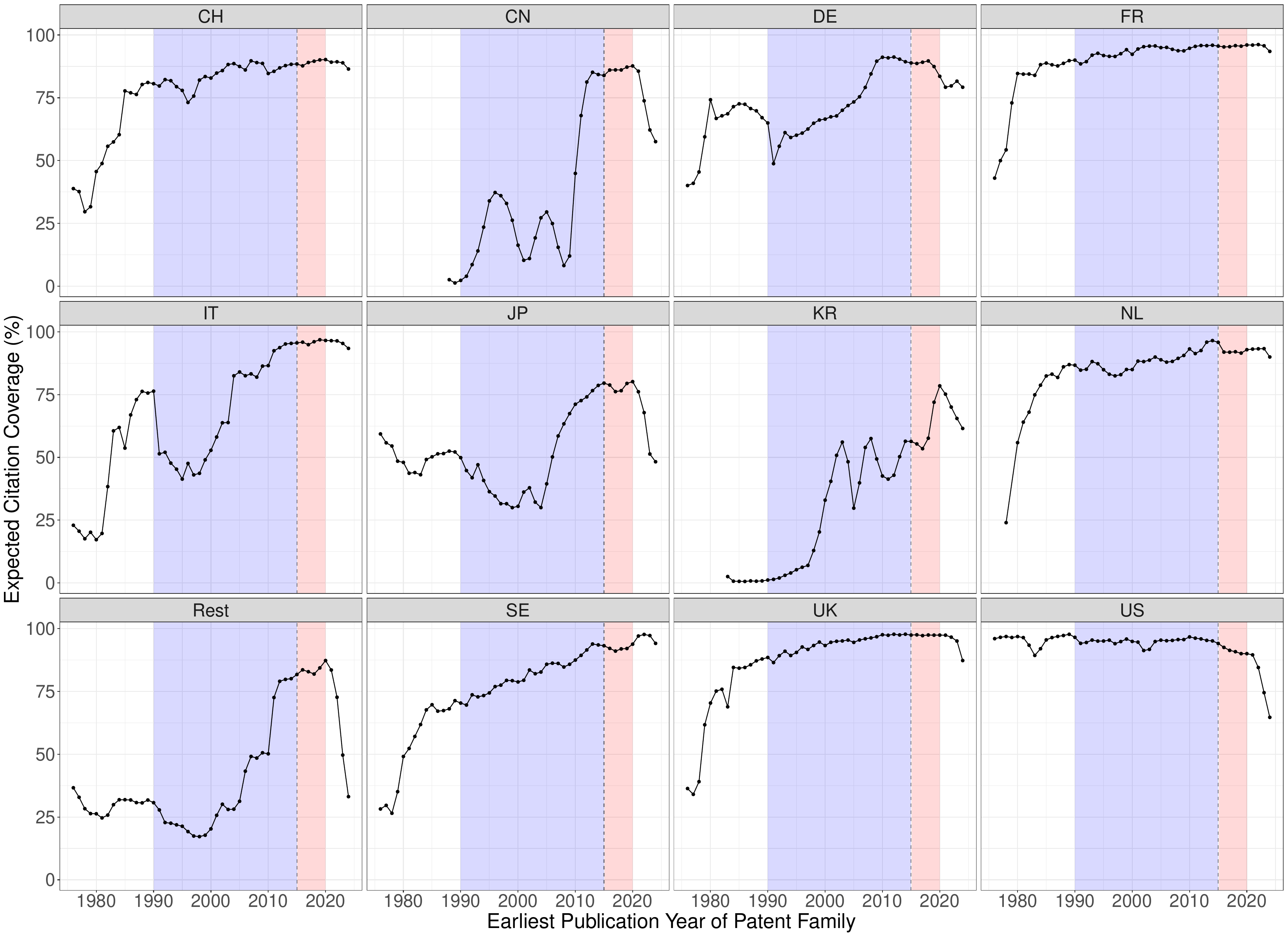}\par\vspace{0.5cm}
\begin{minipage}[b]{\textwidth}
\footnotesize
\emph{Notes:} This figure displays the expected citation coverage for major inventor countries based on our Regional Home Office bottleneck metric. Coverage represents the expected share of patent families in a given country-year cohort for which backward citation data are available in the PATSTAT database. The horizontal axis indicates the earliest publication year of the patent family. To ensure the stability of coverage estimates and eliminate small-sample noise, data points are only plotted for cohorts containing at least $1,000$ unique patent families. The blue-shaded region identifies our primary empirical sample window from $1990$ to $2015$. The red-shaded region marks the five-year forward citation window required for computing the $CD$ index. Beyond this window, right-censoring due to publication delays and database ingestion lags renders backward citation records increasingly incomplete.
\end{minipage}
\end{figure} 

Figure \ref{pat_coverage} plots the resulting expected coverage rate, revealing the share of patent families in each country-year cell for which backward citation data are available in PATSTAT. The unshaded regions document two patterns. (1) data prior to $1990$ exhibit severe missingness and volatility for most European and Asian inventor countries; and (2) data published after $2015$ suffer from right-censoring. Patents citing a focal invention within the five-year forward window must themselves have complete backward citation records for the $CD$ index to correctly distinguish focal-only from combined citations. As publication delays and national examination lags reduce backward citation coverage for recent cohorts, the five-year forward window required for the $CD$ index becomes increasingly unreliable beyond $2020$. We therefore restrict our focal patent sample to the stable publication coverage window between $1990$ and $2015$ (Figure \ref{pat_coverage} blue-shaded region). The corresponding five-year forward citation window extends into the red-shaded region, where backward citation coverage remains adequate but begins to decline. Restricting the focal patent sample to $2015$ helps ensure that even the tail end of the forward window ($2020$) precedes the sharpest coverage deterioration visible. We further control for this citation coverage residual variation within our econometric models.

Table~\ref{tab:sumstats} presents summary statistics for the main variables within this high-fidelity 1990--2015 window.\footnote{Appendix Table~\ref{tab:sumstats_ctry} shows the distribution of DOCDB family-country observations by country of invention. The US has the largest share (45.4\%), followed by Japan (20\%), and Germany (7\%).} The unit of observation is at the DOCDB family-country level. The refined data include 4{,}845{,}084 unique focal patent family observations. Average $N_F$ (forward-only), $N_C$ (combined) and $N_P$ (predecessor-only) forward citations, as well as backward citations, are larger in the extended PATSTAT data than in the restricted USPTO data, which reflects the broader citation coverage of the extended data. The extended PATSTAT data also notably yield a lower average $CD_5$ index than the restricted USPTO data.

\begin{table}[!ht]
\centering
\caption{Summary Statistics at DOCDB Patent Family Level}
\label{tab:sumstats}
\begin{tabular}{lccccc}
\toprule \toprule
Variable & $N$ & Mean & SD & Min & Max \\
  \midrule

$N_{F,0}$ & 4845084 & 4.53 & 8.03 & 0 & 1302 \\
$N_{F,1}$ & 4845084 & 2.97 & 6.40 & 0 & 1289 \\
$N_{C,0}$ & 4845084 & 3.99 & 10.92 & 0 & 1419 \\
$N_{C,1}$ & 4845084 & 3.22 & 10.61 & 0 & 1412 \\
$N_{P,0}$ & 4845084 & 256.80 & 707.74 & 0 & 70370 \\
$N_{P,1}$ & 4845084 & 175.16 & 536.59 & 0 & 58075 \\
$CD_{0}$ & 4845084 & 0.03 & 0.12 & -1 & 1 \\
$CD_{1}$ & 4845084 & 0.05 & 0.19 & -1 & 1 \\
Publication Year & 4845084 & 2005  &  7 & 1990 & 2015 \\
Num Inventors & 4845084 & 2.76 & 2.15 & 1 & 114 \\
Num Inventor Countries & 4845084 & 1.13 & 0.39 & 1 & 9 \\
Num Backward Citations$_{0}$ & 4845084 & 23.24 & 61.38 & 0 & 6548 \\
Num Backward Citations$_{1}$ & 4845084 & 17.43 & 51.87 & 0 & 6210 \\
\bottomrule \bottomrule
\end{tabular}
\begin{tablenotes} 
\footnotesize
\item \textit{Notes:} $N_{F,s}$, $N_{C,s}$, and $N_{P,s}$ respectively denote focal-only, combined, and predecessor-only forward citations as defined in Equation~\eqref{cd_index}. The subscripts indicate the citation data source $s$: $0$ the extended sample (PATSTAT); $1$ the restricted sample (USPTO). The $CD$ index is bounded between $-1$ (fully consolidating) and $1$ (fully disruptive).
\end{tablenotes}
\end{table}

Our Regional Home Office coverage metric and the 1990--2015 sample restriction address the most severe forms of temporal truncation, but two cross-sectional limitations remain. First, Asian patent office citation coverage in PATSTAT is relatively incomplete for certain historical periods even within our sample window (as Figure \ref{pat_coverage} shows). Second, institutional differences present a separate set of challenges. Patent examiner citation practices vary across jurisdictions, examiner- and applicant-added citation distinctions are not recorded uniformly, and varying disclosure requirements may produce systematic self-citation differences. Our baseline restriction of evaluating only USPTO focal patents helps hold the overarching institutional regime constant and mitigates many of these reporting discrepancies across jurisdictions.

\subsection*{Empirical Model: Stacked Regression}

We quantify the measurement bias in the $CD$ index by estimating a linear model that recovers the regime-specific conditional expectations, $m_{c,s}$, as derived in Equation \eqref{eq:relbias_def}. We stack observations for each DOCDB patent family $i$ across both the restricted USPTO ($s=1$) and extended PATSTAT ($s=0$) citation networks. Each patent family therefore appears twice in the stacked dataset---once per citation source. The identifying logic follows the bias estimation literature when multiple measurements of the same underlying observation are available \cite[][]{abowd_stinson2013, bollinger2006}: 

\begin{equation}\label{eq:emp_model} 
CD_{i,c,s,j,t} = \alpha + \sum_{c \neq \mathrm{US}} \beta_c D_{i,c} + \theta R_{i,s} + \sum_{c \neq \mathrm{US}} \delta_c (D_{i,c} \times R_{i,s}) + \mathbf{X}_i'\gamma + \mu_{j,s} + \nu_{t,s} + \varepsilon_{i,c,s,j,t},
\end{equation}

\noindent where $i$ denotes the focal patent family, $c$ the inventor country, $s$ the data source, $j$ the technology field, and $t$ the publication year. $D_{i,c}$ is an inventor country indicator, and $R_{i,s} = \mathbf{1}\{\text{Restricted}\}$ is a source indicator that varies across the stacked observations.

The parameter $\alpha$ identifies the baseline mean disruptiveness of US patents in the extended data ($m_{\text{US},0}$). The coefficient $\beta_c$ quantifies the structural difference in disruptiveness between country $c$ and the US ($m_{c,0} - m_{\text{US},0}$). The parameters of primary interest, $\theta$ and $\delta_c$, map directly to the theoretical measurement bias terms: $\theta$ captures the baseline US bias ($\text{Bias}_{\text{US}} = m_{\text{US},1} - m_{\text{US},0}$); $\delta_c$ represents the relative bias between country $c$ and the US ($\text{RelBias}_{\text{c}} = \text{Bias}_{\text{c}} - \text{Bias}_{\mathrm{US}}$). By estimating these components simultaneously, the model disentangles cross-country variation in innovation performance from citation-network-induced measurement error.

Equation \eqref{eq:emp_model} also includes source-specific technology ($\mu_{j,s}$) and year ($\nu_{t,s}$) fixed effects, which flexibly absorb all common source-specific heterogeneity---including $\alpha$ and $\theta$, which are no longer separately identified. The coefficients $\beta_c$ and $\delta_c$ retain their structural interpretation. To ensure that the estimates are not driven by compositional differences, we include a vector of patent-level controls, $\mathbf{X}_i$, that account for team size and international collaboration. We also include the country-year level mean citation coverage rate interacted with the source indicator to absorb residual variation in PATSTAT data completeness not captured by the 1990--2015 sample restriction. Finally, we control for backward citation counts (measured in the respective sample) to account for the mechanical fact that a larger prior art base increases the probability of subsequent patents referencing those predecessors, which artificially pushes the $CD$ index toward consolidation ($N_C$ and $N_P$). This allows us to isolate the specific forward-citation measurement bias that our framework is designed to identify and correct.

We account for the mechanical correlation resulting from the repetition of patent families across data sources and their distribution across multiple countries and technologies using two-way standard error clustering at the patent-family ($i$) and technology-year ($j,t$) levels. We also report results from a simpler specification that replaces the source-interacted fixed effects with non-interacted technology ($\mu_j$) and year ($\nu_t$) fixed effects, which allows $\theta$ to be separately identified. Appendix Table \ref{tab:main_regression} provides the complete estimation results for all specifications.

\section{Baseline Results} 

\begin{figure}[ht!]\caption{Decomposition of Measurement Bias in Patent Disruptiveness.}\label{avg_disr_rel_us}
\centering\captionsetup{justification=centering}
\includegraphics[width=1\linewidth,keepaspectratio]{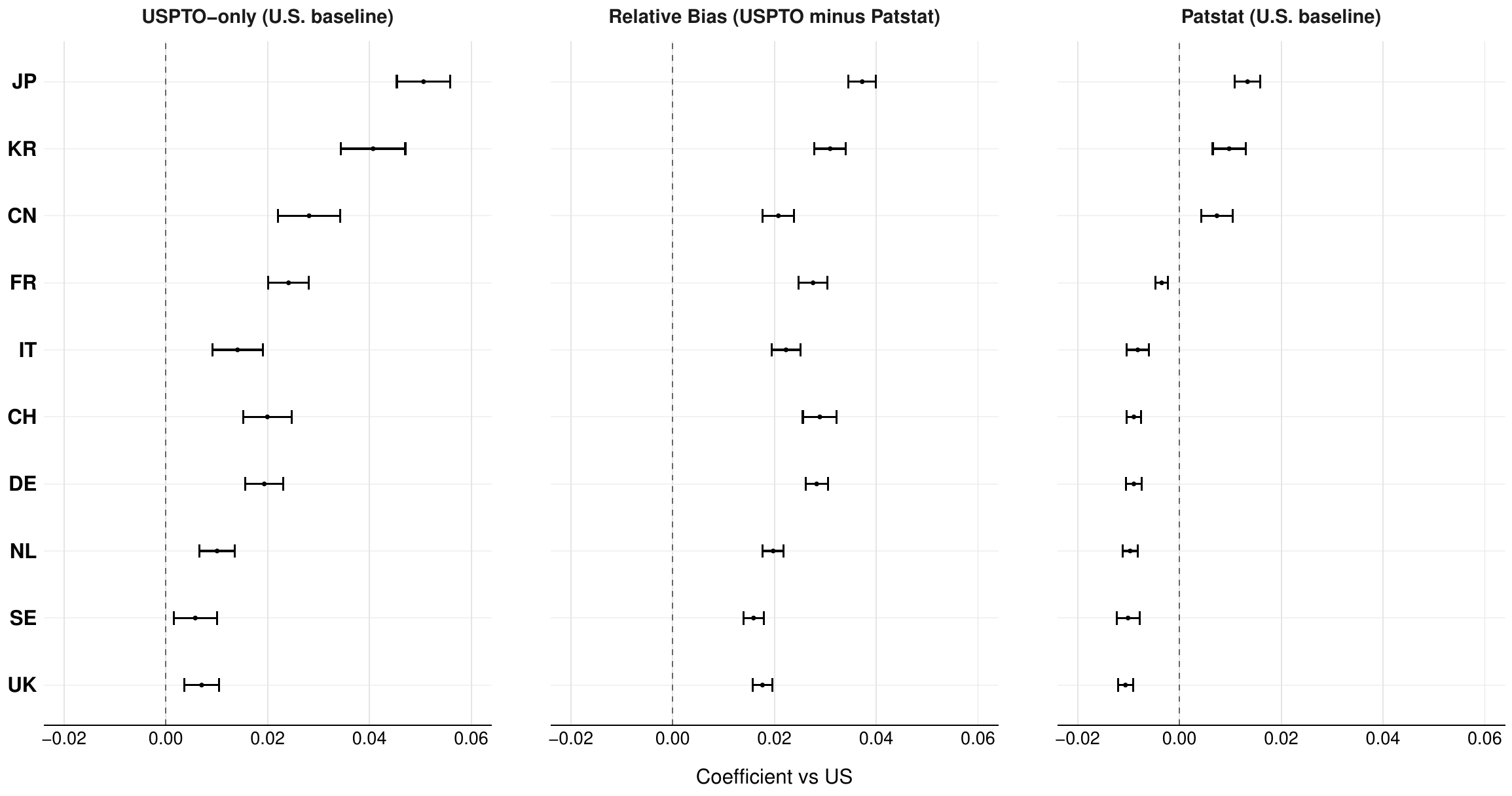}\par\vspace{0.5cm}
\begin{minipage}[b]{\textwidth}
\footnotesize
\emph{Notes:} The figure displays estimated country coefficients from Equation \ref{eq:emp_model}. \textbf{Left Panel:} Estimates using restricted USPTO data ($\beta_{c} + \delta_{c}$), where positive values indicate that non-US inventors appear more disruptive than US inventors. \textbf{Middle Panel:} Relative measurement bias ($\delta_{c}$). Positive values show that USPTO-only data systematically inflates measured disruptiveness for European and Asian countries. \textbf{Right Panel:} Bias-corrected structural estimates ($\beta_{c}$) using extended PATSTAT data. All panels share a common x-axis scale to ensure the bias magnitude (middle) is directly comparable to the baseline (left) and structural signal (right). \textbf{Controls:} All estimates include the fixed effects and patent-level controls described in Section \ref{sec:data}. The US serves as the baseline. Bars represent $95\%$ confidence intervals clustered at the patent family and technology-by-year level.
\end{minipage}
\end{figure}

Figure~\ref{avg_disr_rel_us} presents the fully-interacted specification results of Equation~\ref{eq:emp_model}. This visualization isolates the structural innovation differences ($\beta_c$) from the relative measurement bias ($\delta_c$), providing a corrected global disruptiveness benchmark. The left panel shows country-level differences as measured using the restricted USPTO data ($\beta_c + \delta_c$). Consistent with Figure \ref{fig:intro_uspto_bias}, non-U.S. inventors appear substantially more disruptive than their US counterparts: European country coefficients range between $0.006$ and $0.025$; Asian country coefficients range even higher between $0.028$ and $0.05$. These differences are statistically significant and suggest---at face value---that US inventors lag in disruptive innovation. 

The middle panel isolates the relative measurement bias ($\delta_c$). The baseline US bias is absorbed by the source-specific fixed effects. Using a simpler specification, Column 3 of Appendix Table \ref{tab:main_regression} shows that this baseline bias is negligible and insignificant for the US reference group, representing the second term of equation \eqref{eq:relbias_def}. The relative bias for non-US inventors is consistently large and positive: in particular, estimates range from 0.016 for Sweden to 0.029 for Switzerland, with Germany and France at 0.028. The bias for Japan ($0.037$) and South Korea ($0.031$) exceeds even the largest European estimates, while China ($0.021$) falls within the European range.

To gauge the economic magnitude of this distortion, we benchmark our estimates against the baseline disruptiveness of patents by US inventors in the extended PATSTAT data. Because the constant is absorbed by fixed effects in our main specification, we recover the baseline mean by computing the average predicted value for US patents in the PATSTAT citation network ($s=0$). This value ($\hat{\alpha} = 0.025$) represents a bias-adjusted reference level for disruptiveness. Relative to this benchmark, the measurement bias ranges from 64\% of the US baseline mean for Sweden to 148\% for Japan.\footnote{Percentages are computed as $100 \times \delta_c / \alpha$.} This result confirms that citation truncation in USPTO data does not merely add noise, but rather mechanically and substantially inflates the measured disruptiveness of foreign inventors relative to US inventors.

To further contextualize this magnitude, we compare our estimated bias to a documented determinant of patent disruptiveness. \citet{feng2020crafting} show that examiner assignment variation generates a standard deviation in the $CD$ index of approximately $0.029$. Since our estimated measurement bias for major European countries is of similar magnitude ($\delta_{c} = 0.028$), citation truncation appears to distort patent outcomes at least as significantly as established institutional drivers. This distortion is a systematic feature of the data, rather than an outlier-driven result. Appendix Figure \ref{cumulative_dist} confirms this by showing that the entire distribution of $CD_5$ values for both European and Asian countries shifts leftward when moving from restricted USPTO to extended PATSTAT data. This shift is most pronounced in the positive range where our theoretical framework predicts the largest upward bias.

The right panel reports the corrected cross-country differences based on the extended PATSTAT data ($\beta_c$) using US inventors as the reference group. Once citation truncation is accounted for, the apparent US disadvantage notably disappears: all European country-specific estimates turn significantly negative, ranging from $-0.003$ for France to $-0.011$ for the United Kingdom. These estimates correspond to a measurable structural deficit relative to the US baseline: for instance, the structural deficit of British inventors amounts to 44\% of the US baseline mean.\footnote{Percentage gaps are computed as $100 \times \beta_c / \alpha$.} The pattern differs for the major Asian inventor countries, where structural estimates remain positive even after bias correction: $0.007$ for China, $0.01$ for South Korea, and $0.013$ for Japan. The magnitude of these corrected differences is substantially smaller than the restricted USPTO data suggest, however, indicating that a large share of the apparent Asian advantage is also attributable to measurement bias rather than genuinely higher disruptiveness.

In summary, correcting for citation network truncation fundamentally changes the assessment of disruptive invention leadership. US patents are found more disruptive on average than those from all major European countries and nearly as disruptive as those from major Asian inventor countries once citation network measurement distortion is corrected. These results demonstrate that USPTO-only data produce substantial bias compared to a more complete but still imperfect global citation network. To the extent that PATSTAT itself underreports international citations---particularly for Asian patent offices as documented in Figure \ref{pat_coverage}---our bias estimates represent a lower bound on the true distortion in USPTO-only data. We thus suggest that our estimates of the US structural advantage be viewed as conservative. We do not claim to recover a "true" disruptiveness metric free of all measurement error, however, but instead suggest that our approach provides a more representative view of the global innovation landscape.

\section{Robustness Results\label{sec_robust}}

We evaluate the robustness of our findings by testing sensitivity to citation windows and filing years, restricting the sample to triadic and domestic patents, and examining the stability of the bias across different time periods and technological fields.

\paragraph{$CD$ Index Citation Window Counts and Filing Year}

The baseline analysis examines cross-country patent disruptiveness differences using $CD_{5}$, the standard five-year forward citation window measured from the earliest patent family publication year. We test whether the baseline results vary using a longer ten-year window, $CD_{10}$, or the earliest worldwide filing year as the anchor.

Appendix Figure \ref{avg_disr_rel_us5_10} demonstrates that the uncorrected differences ($\beta_c + \delta_c$), estimated relative bias ($\delta_c$), and bias-corrected structural estimates ($\beta_c$) remain remarkably consistent across all tested models. When expanding to a ten-year window ($CD_{10}$), the bias corrected structural estimates in the extended PATSTAT data shift somewhat. Asian country coefficients become slightly more positive while European coefficients generally become more negative compared to $CD_5$. Using the earliest filing year rather than the publication year as the dating convention yields nearly identical results. Overall, the cross-country ranking remains structurally consistent and these differences do not alter our main qualitative conclusions.

\paragraph{Triadic Patents}

We replicate the analysis using triadic patent families, defined as those granted at the USPTO and filed at the EPO and the Japan Patent Office (JPO). Triadic patents isolate a subset of high-value inventions that are internationally visible and more directly comparable across jurisdictions, thereby reducing asymmetric citation coverage concerns or country-specific filing strategies.

The results in Appendix Figure \ref{p_triadic} provide nuanced validation of our main findings. The relative measurement bias remains sizable and comparable to the full sample; however, the bias corrected structural estimates reveal a more mixed pattern across European countries. Unlike the full sample where European effects are uniformly negative, the triadic sample shows the US advantage holds for major countries like the United Kingdom, whereas coefficients for other countries attenuate toward zero. This lack of a uniform US disruptive advantage in the triadic sample is likely driven by compositional differences. The triadic filter requires filing at the EPO and JPO, which systematically favors internationally-oriented inventions in mature technological fields and under-represents digitally-intensive sectors that constitute a large and growing share of US patenting activity (Appendix Table \ref{tab:composition_bias}). Moreover, the EPO's "technical character" requirement for computer-implemented inventions may act as a legal filter that systematically excludes disruptive US software patents that do not seek or qualify for European protection. The triadic sample therefore compares countries on a systematically different technological and institutional basis than the full sample, which likely accounts for the attenuated US advantage relative to European countries.

The triadic results reinforce the conclusion that the restricted USPTO data induce sizable measurement biases but also importantly show that data commonly used for international comparability introduce their own compositional and legal distortions. In short, triadic patents provide a clean benchmark for comparability, and the full USPTO sample offers a broader and more representative view of disruption at the technological frontier.

\paragraph{Single-Country Inventor Team Restrictions} 

The baseline analysis assigns patent families to inventor countries following standard practice, reflecting the collaborative nature of modern invention. As a robustness check, we restrict the sample to patents with single-country inventors, eliminating multi-country attribution and international team composition. Appendix Figure~\ref{p_single_ctry} is virtually identical to the baseline results: both the magnitude of the structural deficit for European countries and the scale of the relative measurement bias persist. The same holds for the Asian inventor countries, where the positive structural estimates and the large bias terms remain qualitatively unchanged. This confirms that the documented distortions are not driven by cross-border collaboration, but instead reflect fundamental features of the citation network.

\paragraph{Different Time Periods}

We test whether the observed bias changes over time by re-estimating the stacked regressions separately for two sub-periods: 1990–1999 and 2000–2015. Appendix Figure~\ref{avg_disr_rel_us_time} provides the corresponding results. Across both sub-periods, the restricted USPTO data produce the familiar pattern in which non-US countries appear more disruptive than the US. The measurement bias and the structural country effects are broadly stable across sub-periods, with wider confidence intervals in the 1990s reflecting smaller sample sizes particularly for China and South Korea. These results confirm that the documented distortion is not confined to a specific time period but is a persistent feature of the citation network.

\paragraph{Technological Fields}

We assess whether the observed citation bias varies across technologies by re-estimating the stacked regressions separately for six major technology groups: Chemical, Electronics, Instruments, IT, Mechanical, and Pharmaceuticals. Appendix Figure~\ref{avg_disr_rel_us_field} provides the corresponding results. Across all technologies, the restricted USPTO data produce the familiar pattern of inflated disruptiveness among non-US countries, whereas incorporating the extended PATSTAT data largely eliminate these differences. Some fields, notably Chemical and Pharmaceuticals, yield less precise estimates due to smaller sample sizes, but the direction and statistical significance of the bias remain consistent.

\section{Conclusion}

This paper examines cross-country differences in patent disruptiveness using the widely-used $CD$ index. We show that the standard empirical calculation of the index, which relies on USPTO citation data only, creates a structural identification problem for cross-country comparisons because international citations are systematically missing. Our conceptual framework formalizes how this omission distorts the $CD$ index and derives the conditions under which the resulting bias is systematically upward for non-US inventors.

We address this problem by constructing two complementary datasets, both deduplicated at the DOCDB family level to ensure strict comparability. The first follows common practice and uses USPTO data. The second extends the USPTO data with citations from all patent offices covered in PATSTAT data. Using a stacked regression framework that simultaneously identifies structural country differences and source-specific measurement error, three main findings stand out. 

First, country-level disruptiveness measured using the restricted USPTO data is systematically overstated for all non-US inventor countries. The measurement bias inflates non-US disruptiveness by 64 to 148 percent of the US baseline mean, a magnitude comparable to the causal effect of examiner assignment documented by \citet{feng2020crafting}.

Second, correcting for this bias fundamentally reverses the cross-country ranking for Europe. US patents are significantly more disruptive on average than those from every major European country in our sample, with structural deficits ranging from 12 to 44 percent of the US baseline mean. The apparent European advantage documented in USPTO-only data is therefore entirely attributable to measurement distortion.

Third, the bias correction substantially reduces but does not eliminate the measured advantage of major Asian inventor countries relative to US inventors. Because PATSTAT coverage remains lower for Asian offices, these residual positive coefficients likely reflect a mix of genuine structural differences and remaining measurement bias. Consequently, our results provide a conservative estimate of the US relative position, and further data integration would likely reduce Asia’s measured advantage.

More broadly, the requirement of internationally integrated citation data extends to any analysis where units of comparison differ in their exposure to international citation networks. Studies comparing firms, regions, or technological fields that vary in their international patenting intensity may face the same structural identification problem that we document here. Distinguishing between residual measurement bias and genuine structural leadership remains an important challenge, particularly for Asian countries where PATSTAT coverage gaps persist. Progress along this dimension may come from improved access to complete national patent databases or from the development of methods that reliably impute missing cross-office citations.

Finally, our findings offer practical guidance for applied researchers. For studies focused exclusively on US inventors where cross-country comparisons are not the object of interest, the negligible baseline bias we document suggests that USPTO data remain adequate. However, any analysis involving cross-country comparisons or non-US inventor subsamples requires internationally integrated citation data to avoid the distortions documented here. We hope this paper will help give the $CD$ index a more prominent role in the economic analysis of innovation across countries.

\clearpage

\bibliographystyle{elsarticle-harv}
\bibliography{main.bib}

\clearpage

\appendix

\section{Distributional Evidence of Citation Bias}

\begin{figure}[!ht]
\caption{Average Disruptiveness Without Patent Family Correction}
\label{fig:app_single_patent}
\centering
\captionsetup{justification=centering}
\includegraphics[width=0.6\linewidth,keepaspectratio]{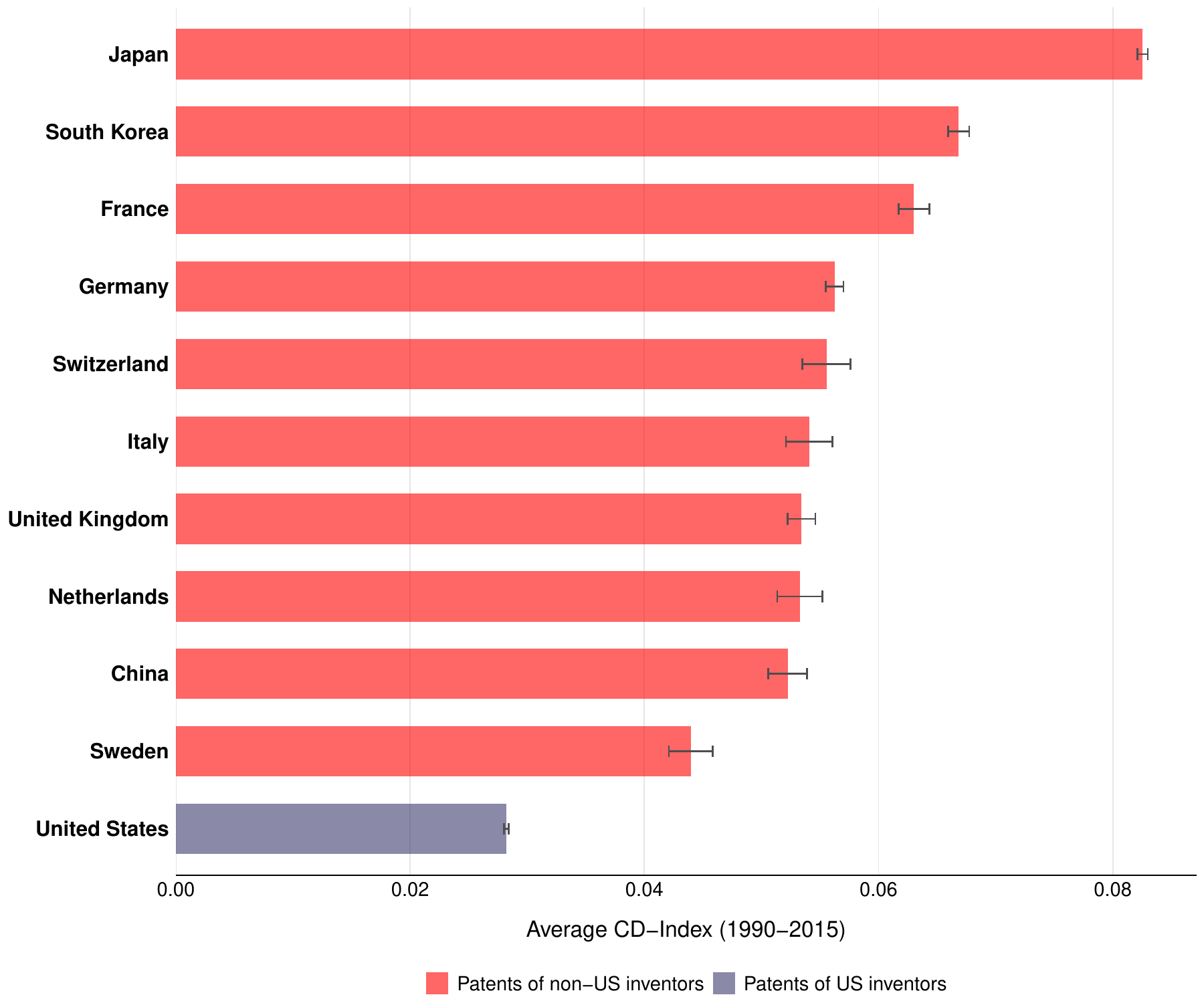}\par\vspace{0.5cm}
\begin{minipage}[b]{\textwidth}
    \emph{Notes:} This figure replicates the analysis in Figure \ref{fig:intro_uspto_bias}, but calculates the $CD$ index at the single patent-level rather than the patent family-level. The sample consists of granted USPTO utility patents (pooled for grant years from 1990 to 2015). The index calculation follows \citet{macher2024} and includes citations to patent applications to avoid the mechanical upward bias in recent years documented in the literature. Error bars represent 95\% confidence intervals. Note that the "US Paradox" (lower measured disruptiveness relative to Europe and East Asia) remains robust and is quantitatively at least as large in this specification, confirming that the bias documented is not driven by the deduplication of patents into families.
\end{minipage}
\end{figure}

\begin{figure}[ht!]
\caption{Cumulative $CD_5$ Distributions by Country and Data Source}
\label{cumulative_dist}
\centering
\captionsetup{justification=centering}
\includegraphics[width=0.9\linewidth,keepaspectratio]{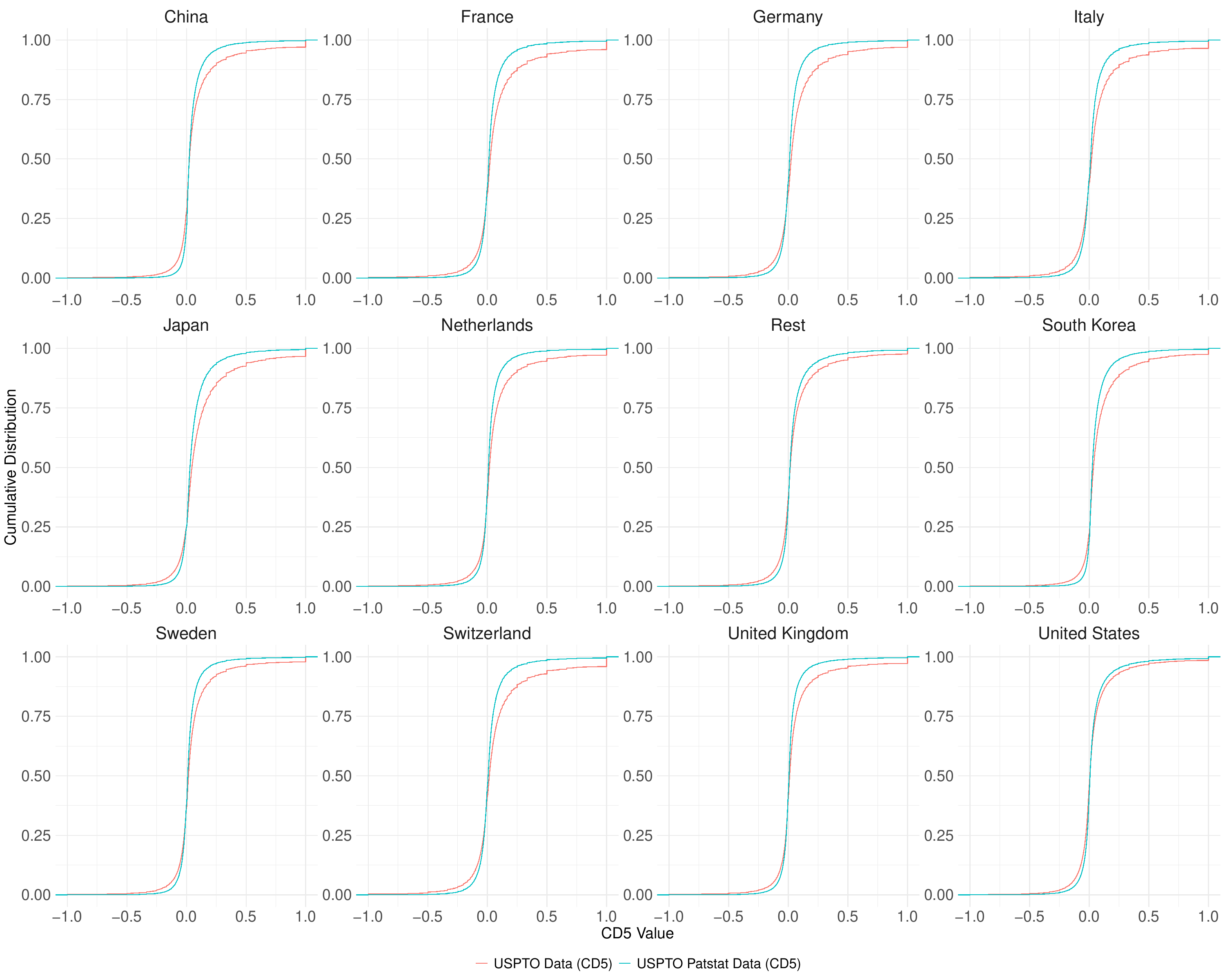}\par\vspace{0.5cm}
\begin{minipage}[b]{\textwidth}
     \emph{Notes:} The figure compares cumulative distribution functions (CDFs) of $CD_5$ values derived from restricted USPTO data (red) and extended PATSTAT data (cyan) for major inventor countries. For the US, the two curves nearly coincide, confirming that adding international citations does not materially alter the disruptiveness distribution for domestic inventors. For European and Asian countries, the PATSTAT-based distribution shifts systematically to the left of the USPTO-based distribution in the positive range, indicating that a large share of patents classified as highly disruptive under the restricted data are revealed to be less disruptive when international backward citations are incorporated. This pattern confirms that the measurement bias documented in the main analysis is structural, affecting the full distribution of positive $CD$ values for non-US\ inventors rather than being driven by outliers.
  
    \end{minipage}
\end{figure}

\clearpage

\section{Empirical Robustness}

\begin{figure}[!ht]
\caption{$CD$ Index Using Alternative Time Windows and Dating Conventions}
\label{avg_disr_rel_us5_10}
\centering
\captionsetup{justification=centering}
\includegraphics[width=0.8\linewidth,keepaspectratio]{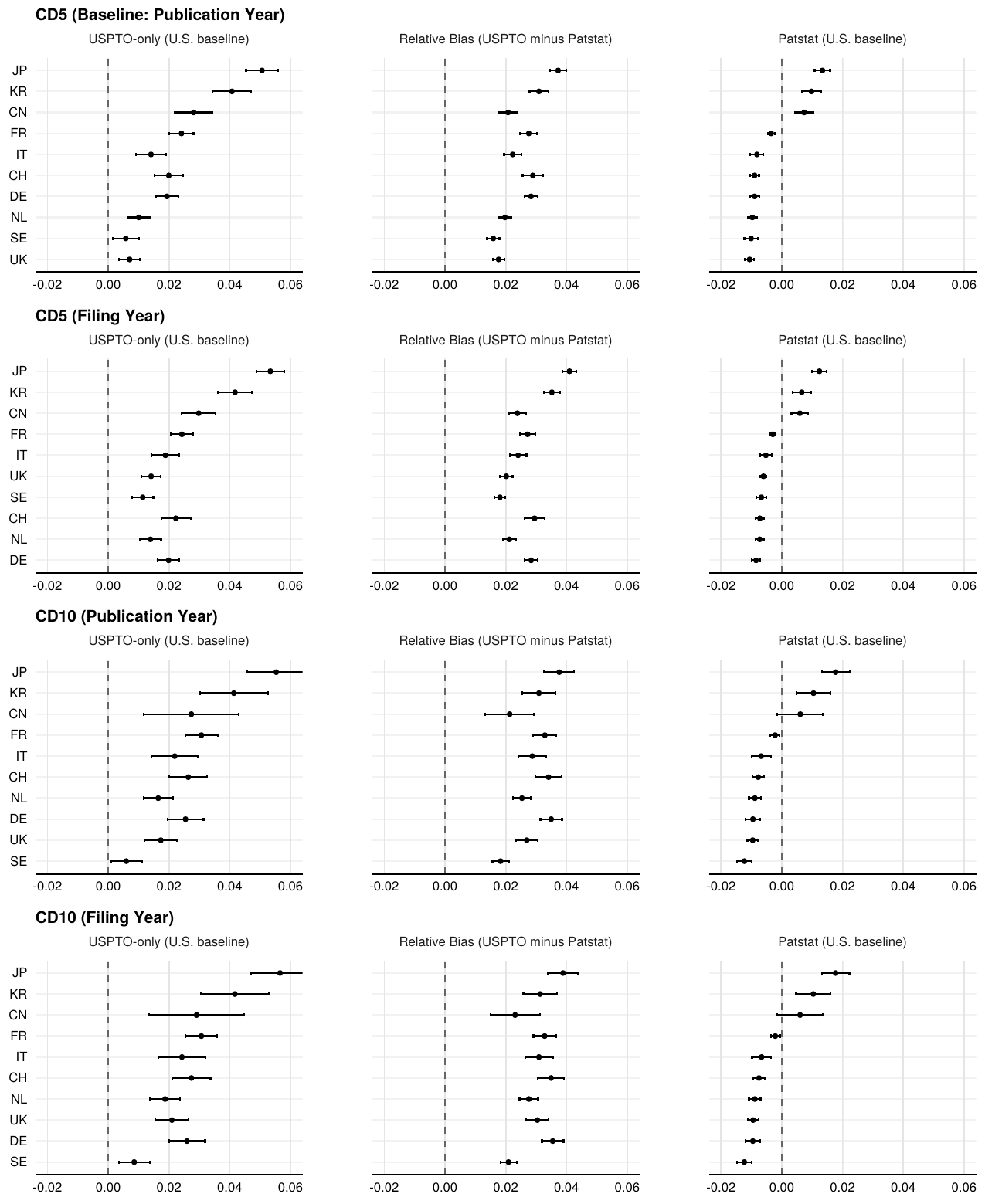}\par\vspace{0.5cm}
\begin{minipage}[b]{\textwidth}
    \emph{Notes:} Stacked regression results comparing cross-country differences in patent disruptiveness using alternative $CD$ index definitions. Each row corresponds to a separate specification defined by the forward citation window ($CD_5$ versus $CD_{10}$) and the patent dating convention (earliest publication year versus earliest filing year). The figures present the decomposition of coefficients relative to the US baseline: the left panels represent the uncorrected apparent difference ($\beta_c + \delta_c$); the middle panels represent the estimated total measurement bias ($\delta_c$); and the right panels represent the bias-corrected structural difference ($\beta_c$). The consistent results across time windows and dating conventions confirm that the baseline findings are robust to these alternative definitions.
    \end{minipage}
\end{figure}

\begin{figure}[ht!]
\caption{$CD$ Index Using Triadic Patent Families}
\label{p_triadic}
\centering
\captionsetup{justification=centering}
\includegraphics[width=1\linewidth,keepaspectratio]{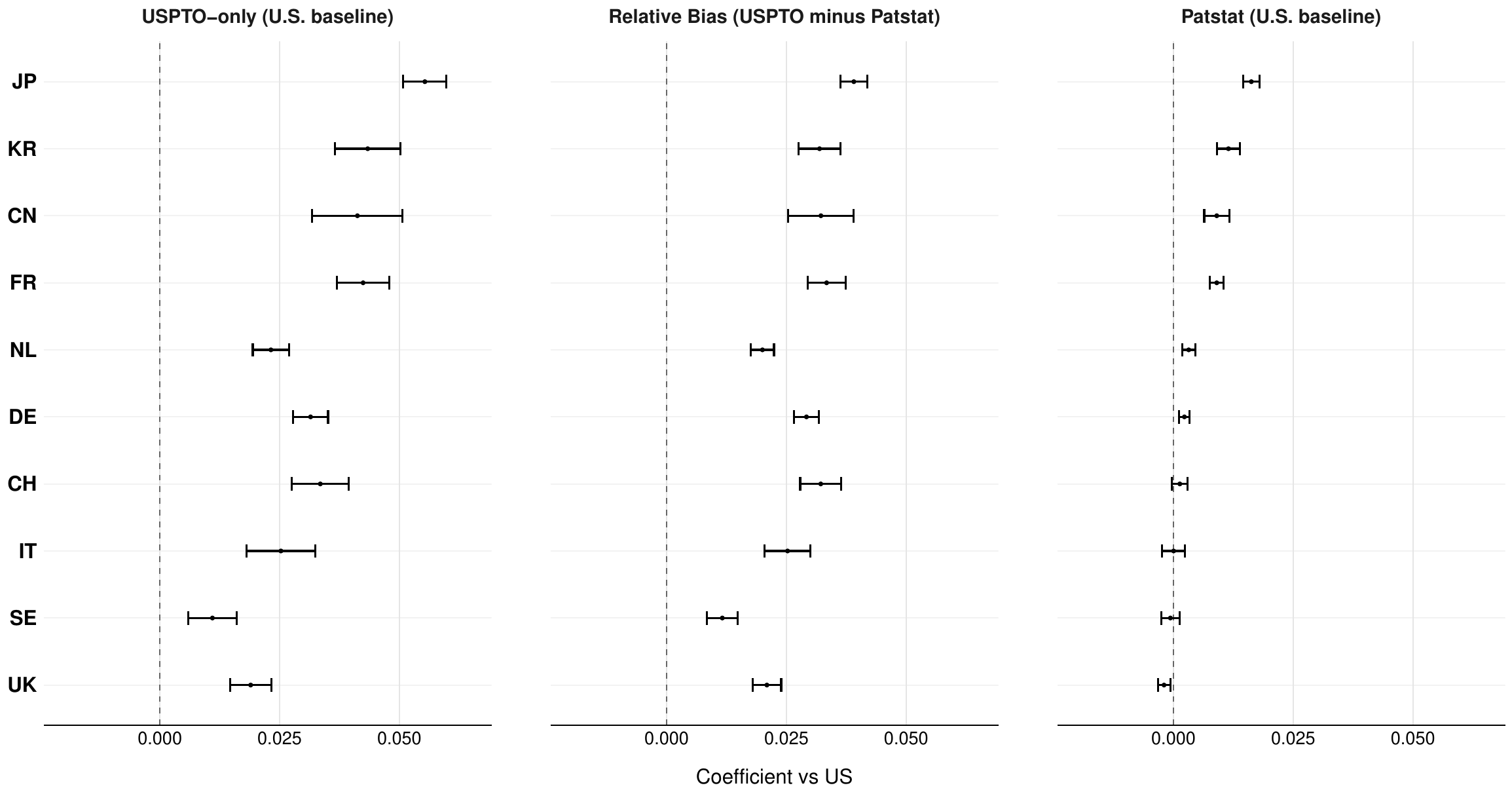} \par\vspace{0.5cm}

\begin{minipage}[b]{\textwidth}
   \emph{Notes:} Stacked regression results restricted to triadic patent families, defined as those granted by the USPTO and also filed at the EPO and the JPO. The left panel shows coefficients using restricted USPTO data (US baseline); the middle panel shows the estimated relative bias ($\delta_c$), representing the difference between USPTO and PATSTAT estimates; and the right panel shows coefficients using extended PATSTAT data ($\beta_c$). The apparent disruptive advantage of major European countries persists for most countries in the right panel. However, the magnitude of the US structural disadvantage ($\beta_c$) is significantly smaller compared to the restricted USPTO-only baseline. As documented in Table \ref{tab:composition_bias}, one primary reason this gap only narrows instead of reversing might be the systematic under-representation of digital patents in the Triadic sample alongside a shift toward more mature technological fields. The resulting comparison is therefore based on a systematically different technological composition than the full sample, which accounts for the shift in the measured coefficients.
   
    \end{minipage}
\end{figure}

\begin{figure}[ht!]
\caption{$CD$ Index Using Single-Country Patent Families}
\label{p_single_ctry}
\centering
\captionsetup{justification=centering}
\includegraphics[width=1\linewidth,keepaspectratio]{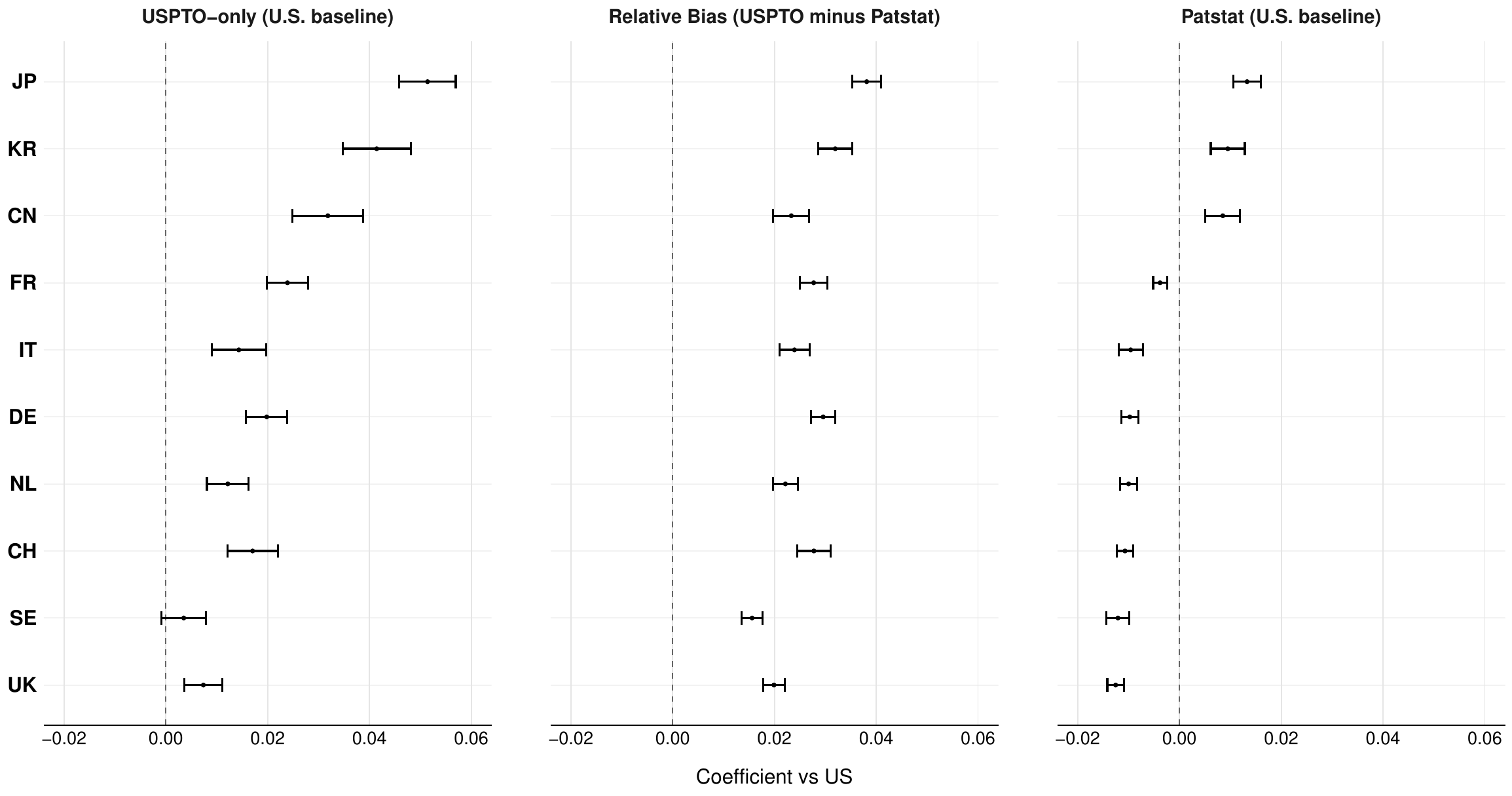} \par\vspace{0.5cm}
\begin{minipage}[b]{\textwidth}
    \emph{Notes:} Stacked regression results restricted to the subsample of single-country patents, defined as those patent families where all listed inventors are located in the same country. The left panel shows coefficients using restricted USPTO data (US baseline); the middle panel shows the estimated relative bias (USPTO minus PATSTAT); and the right panel shows coefficients using extended PATSTAT data. The results are highly consistent with the baseline findings, indicate that the systematic upward bias for foreign inventors persists when excluding multi-country collaborations, and confirm it is not an artifact of the country attribution strategy.
\end{minipage}
\end{figure}

\begin{figure}[ht!]
\caption{$CD$ Index Over Time Periods}
\label{avg_disr_rel_us_time}
\centering
\captionsetup{justification=centering}
\includegraphics[width=1\linewidth,keepaspectratio]{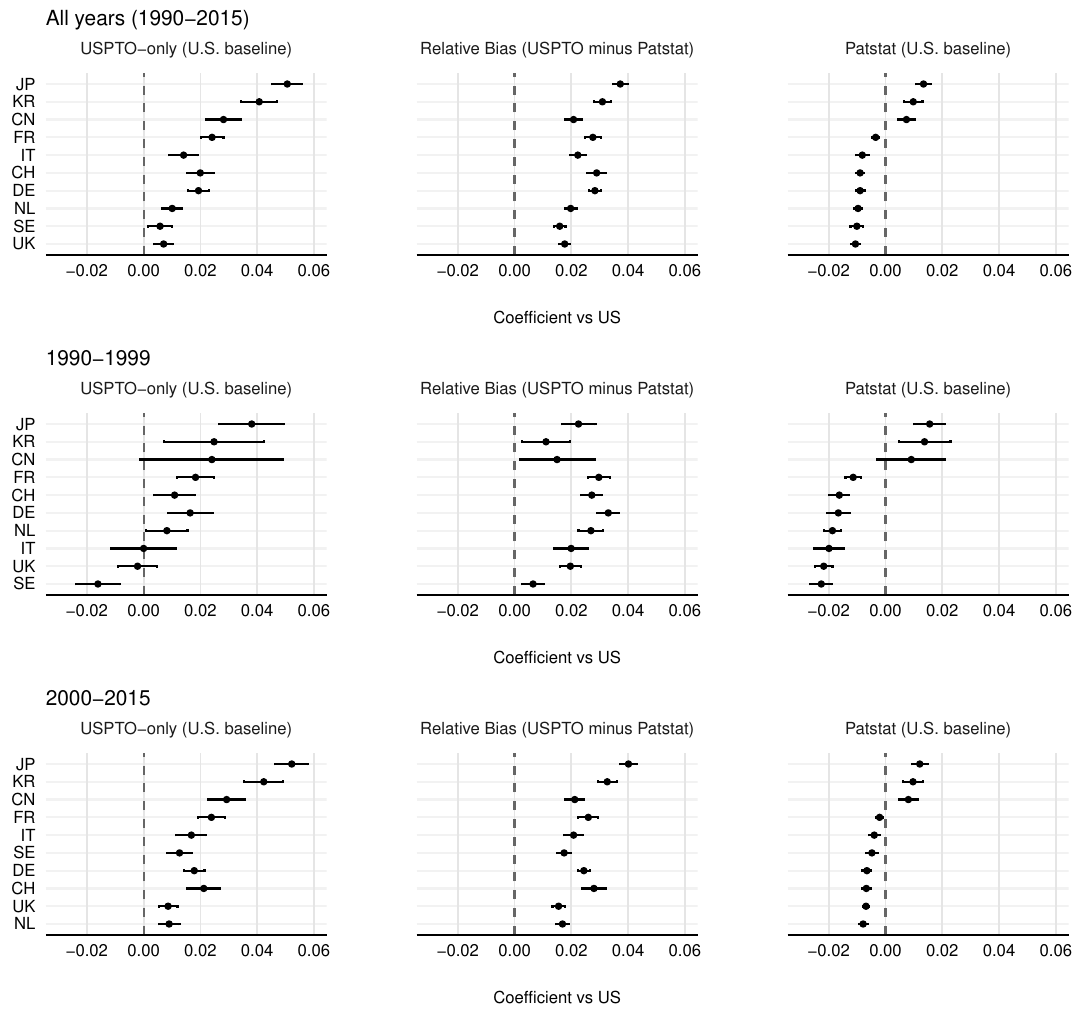} \par\vspace{0.5cm}
\begin{minipage}[b]{\textwidth}
    \emph{Notes:}  Stacked regression results by period: full sample, 1990-1999, and 2000-2015. The left panels show coefficients using restricted USPTO data (US baseline); the middle panels show the estimated relative bias (USPTO minus PATSTAT); and the right panels show coefficients using extended PATSTAT data. The results are highly consistent with the baseline findings, but show the estimated bias declines in the later years.
    \end{minipage}
\end{figure}

\begin{figure}[ht!]
\caption{$CD$ Index Across Technological Fields}
\label{avg_disr_rel_us_field}
\centering
\captionsetup{justification=centering}
\includegraphics[width=\textwidth,keepaspectratio]{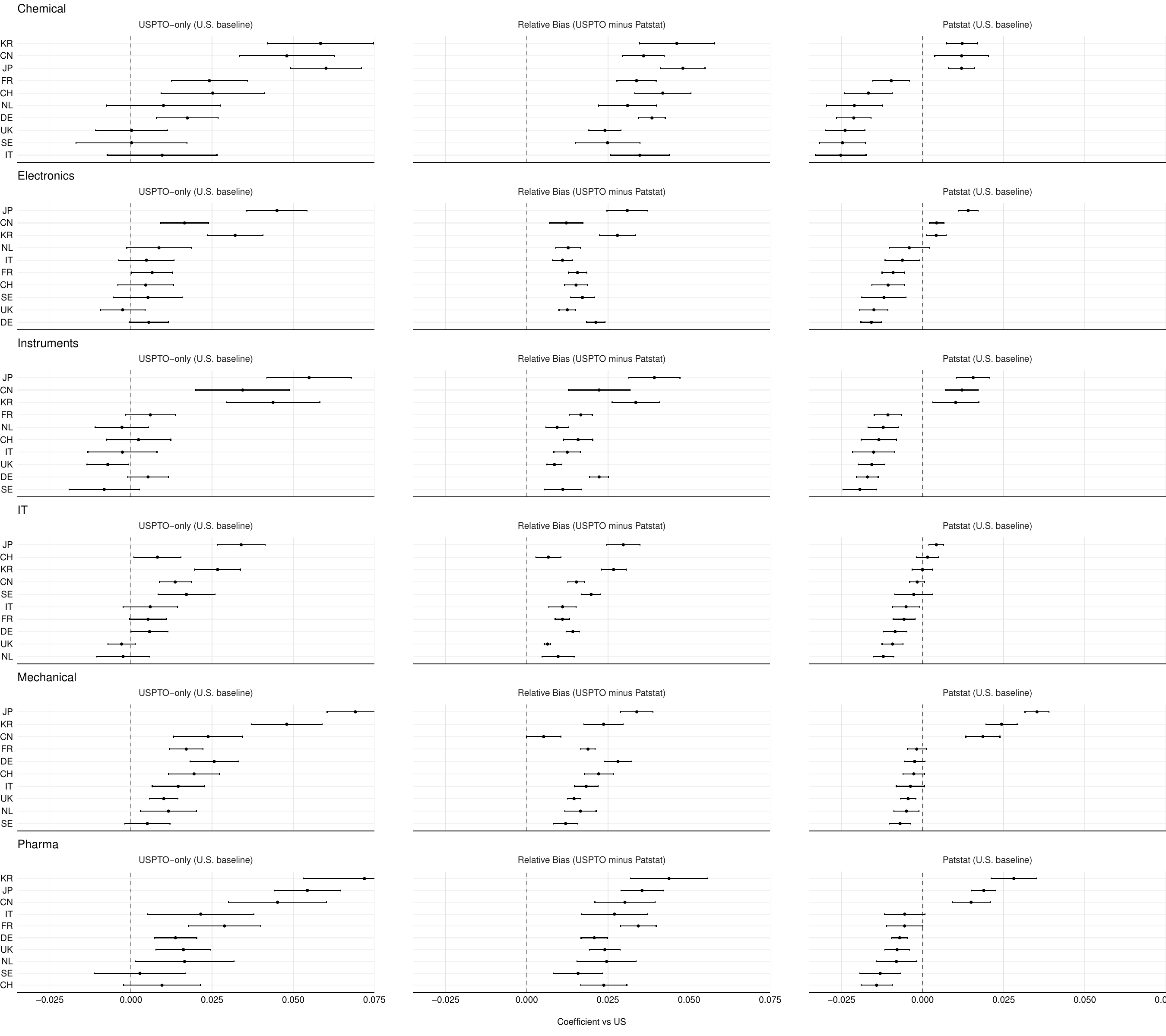}\par\vspace{0.5cm}
\begin{minipage}[b]{\textwidth}
    \emph{Notes:} Stacked regression results by technology group: Chemical, Electronics, Instruments, IT, Mechanical, and Pharmaceuticals). The left panels show coefficients using restricted USPTO data (US baseline); the middle panels show the estimated relative bias (USPTO minus PATSTAT); and the right panels show coefficients using extended PATSTAT data. The results are mostly consistent with the baseline findings, but show some differences in the magnitude of the bias in certain technologies (viz., Chemical and Pharmaceuticals), where smaller sample sizes yield less precise estimates. Some confidence intervals extend beyond the displayed axis range; the direction and statistical significance of all estimates are unaffected.
    \end{minipage}
\end{figure}

\clearpage

\section{Tables}

\begin{table}[ht]
\centering
\caption{DOCDB Family--Country Observation Distribution} 
\label{tab:sumstats_ctry}
\begin{tabular}{lrr}
  \toprule \toprule
  Country & Number & Share \\
  \midrule
  US & 2201557 & 0.454 \\ 
  JP & 967517 & 0.200 \\ 
  DE & 338765 & 0.070 \\ 
  KR & 212341 & 0.044 \\ 
  FR & 133299 & 0.028 \\ 
  UK & 125840 & 0.026 \\ 
  CN & 93083 & 0.019 \\ 
  CH & 52378 & 0.011 \\ 
  NL & 50538 & 0.010 \\ 
  IT & 50104 & 0.010 \\ 
  SE & 46634 & 0.010 \\ 
  Rest & 573028 & 0.118 \\ 
  \bottomrule \bottomrule
\end{tabular}
\begin{tablenotes} 
\item \textit{Notes:} Observations defined at the DOCDB patent family-country level, such that a patent family with inventors from multiple countries contributes one observation to each inventor country. Shares are computed relative to the total number of family-country observations. The \textit{Rest} category includes all other countries, such as Canada, Taiwan, or Spain.
\end{tablenotes}
\end{table}

\begin{table}[ht]
\centering
\caption{Technological Composition: Decadal Averages (1990--2015)}
\label{tab:composition_bias}
\begin{tabular}{llcc}
\toprule
Decade & NBER Field & Full Sample (\%) & Triadic Sample (\%) \\
\midrule
1990s & Chemical & 14.30 & 28.75 \\
1990s & Electronics & 17.96 & 17.24 \\
1990s & Instruments & 12.38 & 11.18 \\
1990s & IT & 11.23 & 10.24 \\
1990s & Mechanical & 27.09 & 19.77 \\
1990s & Pharma & 5.97 & 9.93 \\
1990s & other & 11.07 & 2.89 \\
\midrule
2000s & Chemical & 9.93 & 21.86 \\
2000s & Electronics & 22.22 & 18.53 \\
2000s & Instruments & 12.24 & 10.88 \\
2000s & IT & 25.25 & 15.82 \\
2000s & Mechanical & 18.08 & 17.26 \\
2000s & Pharma & 5.15 & 12.84 \\
2000s & other & 7.15 & 2.81 \\
\midrule
2010s & Chemical & 9.70 & 20.74 \\
2010s & Electronics & 21.17 & 17.75 \\
2010s & Instruments & 11.68 & 10.28 \\
2010s & IT & 29.77 & 17.48 \\
2010s & Mechanical & 16.20 & 16.69 \\
2010s & Pharma & 5.51 & 13.78 \\
2010s & other & 5.96 & 3.28 \\
\bottomrule
\end{tabular}
\begin{tablenotes} 
\item \textit{Notes:} This table presents decadal averages of percentage shares for the full USPTO sample versus the restricted Triadic sample starting in 1990. The results provide empirical evidence of a systematic compositional bias. The share of IT in the full sample grew significantly over time to nearly 30\% by the 2010s, but this technology remains consistently under-represented in the Triadic sample. Mature sectors such as Chemicals are over-represented in the Triadic sample relative to the full sample, as seen in the 1990s where the share is approximately double. 
\end{tablenotes}
\end{table}

\begin{table}[htbp]
\renewcommand{\arraystretch}{0.8}
   \caption{\label{tab:main_regression} Decomposition of Measurement Bias and Structural Country Effects}
   \centering
   \tiny
   \begin{tabular}{lccc}
      \toprule \toprule
      Dependent Variable: & \multicolumn{3}{c}{$CD_5$}\\
       & Interacted (no controls) & Interacted (Main) & Simple FE \\   
      Model: & (1) & (2) & (3)\\  
      \midrule
      
      \multicolumn{4}{l}{\textbf{Measurement Bias Estimates ($\delta_c$)}} \\
      \multicolumn{4}{l}{\textit{(Bias Coefficients: USPTO Data)}} \\ 
      \hspace{1em} Switzerland & 0.029$^{***}$ & 0.029$^{***}$ & 0.031$^{***}$\\   
       & (0.002) & (0.002) & (0.002)\\   
      \hspace{1em} Germany & 0.028$^{***}$ & 0.028$^{***}$ & 0.028$^{***}$\\   
       & (0.001) & (0.001) & (0.001)\\   
      \hspace{1em} France & 0.028$^{***}$ & 0.028$^{***}$ & 0.029$^{***}$\\   
       & (0.001) & (0.001) & (0.002)\\   
      \hspace{1em} United Kingdom & 0.018$^{***}$ & 0.018$^{***}$ & 0.018$^{***}$\\   
       & (0.0010) & (0.0010) & (0.001)\\   
      \hspace{1em} Italy & 0.021$^{***}$ & 0.022$^{***}$ & 0.022$^{***}$\\   
       & (0.001) & (0.001) & (0.002)\\   
      \hspace{1em} Netherlands & 0.020$^{***}$ & 0.020$^{***}$ & 0.021$^{***}$\\   
       & (0.001) & (0.001) & (0.001)\\   
      \hspace{1em} Sweden & 0.016$^{***}$ & 0.016$^{***}$ & 0.015$^{***}$\\   
       & (0.001) & (0.001) & (0.001)\\   
      \hspace{1em} China & 0.019$^{***}$ & 0.021$^{***}$ & 0.015$^{***}$\\   
       & (0.002) & (0.002) & (0.002)\\   
      \hspace{1em} Japan & 0.036$^{***}$ & 0.037$^{***}$ & 0.034$^{***}$\\   
       & (0.001) & (0.001) & (0.001)\\   
      \hspace{1em} South Korea & 0.029$^{***}$ & 0.031$^{***}$ & 0.024$^{***}$\\   
       & (0.002) & (0.002) & (0.002)\\   
      \hspace{1em} Rest of World & 0.013$^{***}$ & 0.015$^{***}$ & 0.011$^{***}$\\   
       & (0.001) & (0.001) & (0.001)\\   
       \addlinespace[0.5em]
      \hspace{1em} US Baseline Bias ($\delta_{US}$) & Absorbed & Absorbed & $-$0.003\\   
       &  &  & (0.003)\\   
      
      \addlinespace[1em]
      
      \multicolumn{4}{l}{\textbf{Structural Country Effects ($\beta_c$)}} \\
      \multicolumn{4}{l}{\textit{(Main Effects: PATSTAT Data)}} \\
      \hspace{1em} Switzerland & $-$0.006$^{***}$ & $-$0.009$^{***}$ & $-$0.010$^{***}$\\   
       & (0.0007) & (0.0007) & (0.0008)\\   
      \hspace{1em} Germany & $-$0.006$^{***}$ & $-$0.009$^{***}$ & $-$0.008$^{***}$\\   
       & (0.0007) & (0.0008) & (0.0009)\\   
      \hspace{1em} France & $-$0.0007 & $-$0.003$^{***}$ & $-$0.004$^{***}$\\   
       & (0.0006) & (0.0006) & (0.0007)\\   
      \hspace{1em} United Kingdom & $-$0.010$^{***}$ & $-$0.011$^{***}$ & $-$0.011$^{***}$\\   
       & (0.0007) & (0.0007) & (0.0008)\\   
      \hspace{1em} Italy & $-$0.005$^{***}$ & $-$0.008$^{***}$ & $-$0.007$^{***}$\\   
       & (0.001) & (0.001) & (0.001)\\   
      \hspace{1em} Netherlands & $-$0.008$^{***}$ & $-$0.010$^{***}$ & $-$0.010$^{***}$\\   
       & (0.0007) & (0.0007) & (0.0007)\\   
      \hspace{1em} Sweden & $-$0.007$^{***}$ & $-$0.010$^{***}$ & $-$0.009$^{***}$\\   
       & (0.001) & (0.001) & (0.001)\\   
      \hspace{1em} China & 0.012$^{***}$ & 0.007$^{***}$ & 0.011$^{***}$\\   
       & (0.002) & (0.002) & (0.002)\\   
      \hspace{1em} Japan & 0.018$^{***}$ & 0.013$^{***}$ & 0.016$^{***}$\\   
       & (0.001) & (0.001) & (0.002)\\   
      \hspace{1em} South Korea & 0.015$^{***}$ & 0.010$^{***}$ & 0.015$^{***}$\\   
       & (0.002) & (0.002) & (0.002)\\   
      \hspace{1em} Rest of World & 0.005$^{***}$ & 0.001 & 0.005$^{***}$\\   
       & (0.0009) & (0.0009) & (0.001)\\   
      
      \midrule
      \emph{Controls}\\
      Citation Coverage $\times$ Source & $-$0.006$^{***}$ & $-$0.011$^{***}$ & $-$0.005$^{*}$\\   
       & (0.002) & (0.002) & (0.003)\\   
      Number of Countries & No & 0.001$^{**}$ & 0.001$^{**}$\\   
       &  & (0.0004) & (0.0004)\\   
      Number of Inventors & No & $-$0.107$^{***}$ & $-$0.107$^{***}$\\   
       &  & (0.007) & (0.007)\\   
      Backward Citations & No & $-$0.018$^{***}$ & $-$0.018$^{***}$\\   
       &  & (0.002) & (0.002)\\   
       
      \midrule
      \emph{Fixed Effects}\\
      Source $\times$ Year ($\nu_{t,s}$) & Yes & Yes & No\\  
      Source $\times$ Tech ($\mu_{j,s}$) & Yes & Yes & No\\  
      Year ($\nu_t$) & (Absorbed) & (Absorbed) & Yes\\  
      Tech ($\mu_j$) & (Absorbed) & (Absorbed) & Yes\\  
      \midrule
      \emph{Fit statistics}\\
      Observations & 9,690,168 & 9,690,168 & 9,690,168\\  
      R$^2$ & 0.025 & 0.029 & 0.029\\  
      Within R$^2$ & 0.010 & 0.014 & 0.017\\  
      \bottomrule \bottomrule
\end{tabular}
   \begin{tablenotes} 
\footnotesize
\item \textit{Notes:} Standard errors, reported in parentheses, are two-way clustered at the patent-family ($i$) and technology-year ($j,t$) levels. The reference category for all country estimates is the United States. Columns (1) and (2) employ a fully interacted fixed effects specification, where technology ($\mu_{j,s}$) and year ($\nu_{t,s}$) intercepts are allowed to vary by citation source ($s$). These source-specific fixed effects absorb the baseline source indicator ($\theta$). Column (2) includes patent-level controls for team size (rescaled by 100 for readability) and international collaboration. Column (3) utilizes a simple fixed effects specification ($\mu_j, \nu_t$) to identify the baseline U.S. bias ($\delta_{US}$). Significance levels: *** $p<0.01$, ** $p<0.05$, * $p<0.1$.
\end{tablenotes}
\end{table}

\end{document}